\documentclass[12pt,preprint]{aastex}

\shorttitle{\object{RV UMa}: quintuplets in the frequency spectrum}
\shortauthors{Hurta et al.}

\begin{document}

\title{First quintuplet frequency solution of a Blazhko variable:\\light curve analysis of \object{RV UMa}}

\author{Hurta Zs.\altaffilmark{1}}
\affil{E\"otv\"os Lor\'and University, Department of Astronomy, P.O. Box 32, H-1518 Budapest, Hungary}
\email{hurta@konkoly.hu}

\author{Jurcsik J., Szeidl B., and S\'odor, \'A}
\affil{Konkoly Observatory of the Hungarian Academy of Sciences, P.O. Box 67, H-1525 Budapest, Hungary}
\email{jurcsik,szeidl,sodor@konkoly.hu}

\altaffiltext{1}{Visiting Astronomer, Konkoly Observatory of the Hungarian Academy of Sciences}

\begin{abstract}
\object{RV UMa} is one of the RRab stars showing regular large amplitude light curve modulation. Extended photoelectric observations of \object{RV UMa} obtained at the Konkoly Observatory were published by \citet{ka76}. The analysis of the data was published by \citet{kv95}. After detecting an error
in the reduction procedure of the published Konkoly data, corrected photometric data are presented with additional, previously unpublished measurements.

The reanalysis  of the combination of the corrected Konkoly data
supplemented with  Preston \& Spinrad's (1967) observations has led to the discovery that the adequate mathematical model of the light curve is, in fact, a quintuplet, instead of a triplet frequency solution. This finding has crucial importance in the interpretation of the Blazhko phenomenon, as triplet (doublet) is the preferred structure in the resonance models, quintuplet in the magnetic models.

Period changes of both the pulsation and the modulation light variations of \object{RV UMa} have been detected
based on its century long photometric observations.
An overall anticorrelation between the pulsation and the modulation period changes can be defined 
with $dP_{Bl}$/$dP_0$=$-8.6 \times 10^4$ gradient, i.e., the modulation period is longer if the
pulsation period is shorter. Between 1946 and 1975 the pulsation and modulation periods showed, however, parallel changes, which points to that there is no strict relation between the changes in the periods of the pulsation and the modulation.
\end{abstract}


\section{Introduction\label{int}}

Most RR Lyrae stars repeat their light curve with remarkable regularity. A group of them (about $20-30\%$ of the known galactic RRab stars), however, shows cyclic modulation in the shape and amplitude of their light curves over tens to thousands of pulsation cycles \citep{sm81, sm95, sz88}. Although the effect (called Blazhko effect) has been known for one hundred years, it lacks widely accepted theoretical understanding. The explanation of this phenomenon is a great challenge for the theory and it would undeniably be extremely important for understanding the nature of the pulsation of these stars. Although several ideas have been put forward to explain the modulation behavior e.g., by \citet{dz04,sh00,st06}, none of them can be reconciled with most of the observational facts. At present we can only say that neither the physical nor the exact mathematical model of the Blazhko modulation is known. 

\cite{kv95} writes that ``{\it long-term accurate photometry is needed to establish a mathematical model of the modulation of the light variation, which is a prerequisite of the construction of a physical model}''. In the frequency spectrum of Blazhko stars the triplet structure around the pulsation frequency and its harmonics is a typical feature. From a theoretical point of view it is an important question whether the frequency structure is indeed a triplet structure, or whether the noise of observations and/or the unsatisfactory data sampling make the finer (e.g. quintuplet) structure unobservable, as triplet (doublet) is the preferred structure in the resonance models, quintuplet in the magnetic models. \cite{kv95} attempted to construct a mathematical model for the light variation of \object[RV UMa]{RV Ursae Majoris}, perhaps the most regular Blazhko RRab star, using only a part (seven years) of the Konkoly observations. However, it turned out that the published data he used were erroneous \citep{hu07}.

Among the modulated RRab stars, \object{RV UMa} has the longest, homogeneous photoelectric observations suitable for constructing a mathematical model. The three-color ($UBV$) photometry of the star was started at the Konkoly Observatory in 1958 and spanned over 17 years. In this paper we discuss the corrected and extended Konkoly photometry supplemented with the observations of \citet{pr67}, and new important results are given. In Section \ref{ana}, we give an improved mathematical model of the modulation. Through the study of all published photometric data we also investigate the relation between the changes of the pulsation and modulation periods in Section \ref{ltc}.

\section{Observations\label{obs}}
The published Konkoly photoelectric observations of \object{RV UMa}, which were  
erroneously transformed into the standard system, have been corrected and 
previously unpublished observations between the years 1959 and 1975 have been added \citep{hu07}. 
The new photometry differs from the data used by  \citet{kv95} by $0.00-0.13$ mag, the differences are not constant, so the amplitudes of the light curve in Kov\'acs' discussion are also in error. The photometric accuracy of the individual $B$ and $V$ magnitudes are estimated to be $0.02-0.03$ mag. The complete, corrected photoelectric $UBV$ observations obtained at the Konkoly Observatory 
are available at {http://www.konkoly.hu/24/publications/rvuma}.

As the corrections of the data affect the results published by \citet{kv95}, we reanalyze the entire, corrected Konkoly $B$, $V$ photoelectric data sets, supplemented with \citeauthor{pr67}'s \citeyearpar{pr67} (hereafter P\&S) observations. The latter observations show an extremely large amplitude of the pulsation on JD 2,436,733 which has never been detected at any other epoch. Therefore, these data points were omitted from the analysis. Table \ref{lot} lists the log of observations used in the present analysis.

Although there were no simultaneous Konkoly and P\&S measurements
 consecutive pulsation maxima were observed at the two sites on JD 2,436,724.
Taking into account the regularity and the relatively long modulation period of RV UMa
the amplitude of consecutive pulsation cycles can change only marginally. 
Maximum lights observed at 2,436,724.363 (Konkoly)
and at 2,436,724.835 (Berkeley) are identical within 0.01 mag, and 0.02 mag in $V$ and $B$ bands,
respectively, indicating that there is no significant discrepancy
between the color systems of the corrected Konkoly and the P\&S data.
The published (Kanyo, 1976) maximum brightnesses for
this night were, however, 0.1 mag and 0.05 mag brighter in V and B than the P\&S data.

In Section \ref{ltc}, the long-term changes of the pulsation and modulation properties of \object{RV UMa} are discussed using observations spanning over 90 years that were collected and homogenized by \citet{hu07}.

\section{Light curve analysis of RV UMa\label{ana}}

The aim of the present analysis is to check how accurately the light curve of a Blazhko star showing large amplitude modulation can be modeled with the generally accepted triplet frequency solution.
This mathematical model of the light curve 
involves only 2 independent frequency components, namely the fundamental mode pulsation frequency $f_0$, and the modulation frequency $f_m$. All the other frequencies are linear combinations of these frequencies and can be given as a series of $kf_0 \pm jf_m$, $k=0...n$, $j=0,1$ frequencies. The available extended photometric observations 
of large modulation amplitude Blazhko stars e.g., \object{RR Lyrae} \citep{sm03}, \object{AH Cam} \citep{sm94}, \object{XZ Cyg} \citep{la04} show, besides the modulation,  some kind of irregular
behavior as well, caused by long-term cyclic changes and/or the multiperiodicity of the modulation. The light curve changes of \object{RV UMa} are, however, very stable, giving a unique opportunity to check the validity of modeling the light curve by  equidistant triplet frequency series.

The photoelectric observations utilized cover 17 years. During this interval small but non-negligible
period changes both in the pulsation and the modulation periods occurred 
as the $O-C$ plots and direct period determinations indicate 
\citep{hu07}. These period changes may affect the results of the analysis 
as they may lead to the appearance of spurious signals. Therefore data analysis is performed not only on the whole data set but also on its most densely sampled section (between JD 2,436,647--2,437,463, data from 1959-1961).
Results are presented for the entire $B$ and $V$ data sets ($B1$, $V1$), and their subsamples ($B2$, $V2$), separately.\footnote
{Data sets are available at {http://www.konkoly.hu/24/publications/rvuam}} In Table \ref{fqt} the time interval and number of data points for the 4 data sets are summarized.  The frequencies and their amplitudes and phases have been determined using the facilities of the program package MUFRAN \citep{ko90} and a linear combination fitting program developed by \'Ad\'am S\'odor.

\begin{figure}
\epsscale{.80}
\includegraphics[angle=0,width=16cm]{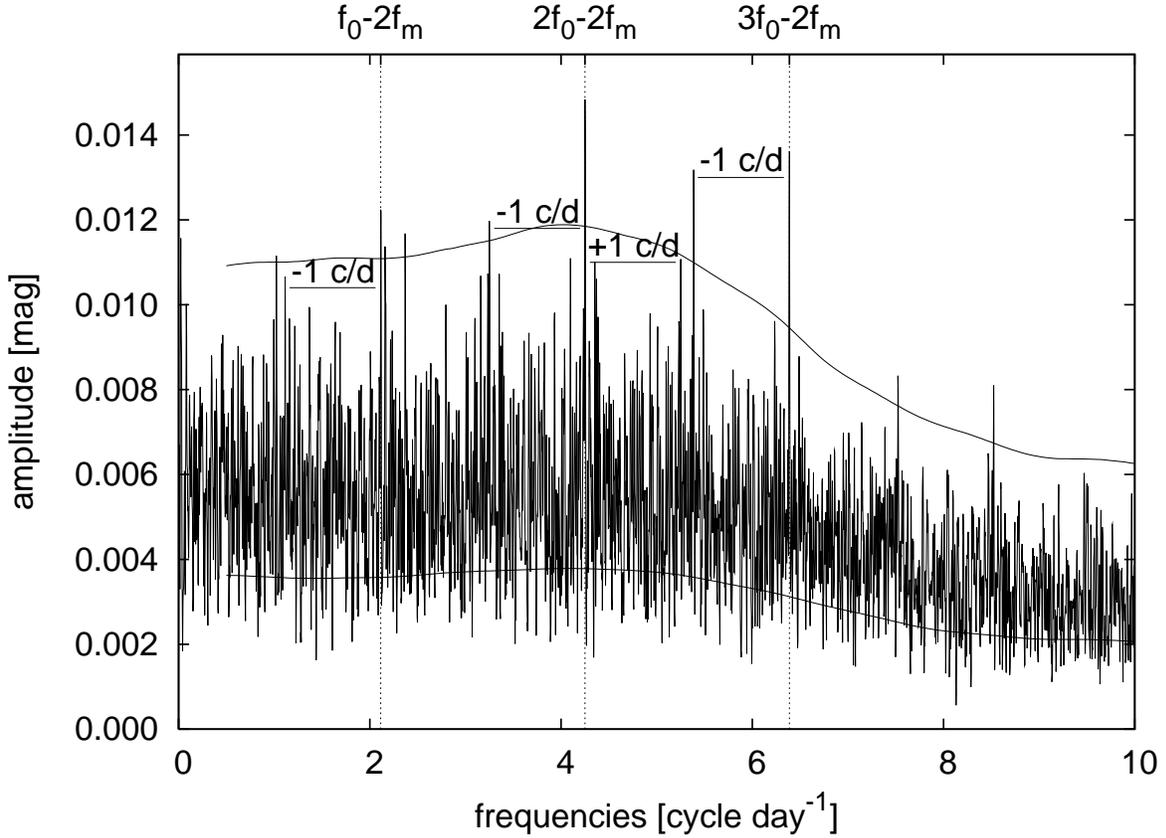}
\caption{Residual spectrum of the $V2$ data set after the removal of the equidistant triplet frequency solution from the light curve. The mean and the 4$\sigma$ limits are also drawn.
Significant signals above the noise level appear at $kf_0 - 2f_m$ ($k=1,2,3$) frequencies
with $4.6$, $5.5$ and $6.2$ $S/N$ ratio. The other higher peaks are $\pm 1$ cycle day$^{-1}$ aliases of these components. RV UMa is the first Blazhko variable with quintuplet frequencies detected in the Fourier spectrum of the light curve. Compressed plotting is applied as described by \citet[Fig~1]{kv95}. \label{v2res1}}
\end{figure}

The data analysis concentrates on the residual spectra after the removal of the supposed equidistant triplet frequencies. 
It is found that the residual spectra show signals above the noise level
at $kf_0 - 2f_m$ frequencies. As an example, the residual spectrum of the $V2$ data set is shown in Figure \ref{v2res1}. The highest peak in this spectrum is at $2f_0 - 2f_m$ frequency. The three highest amplitude signals in 
Figure \ref{v2res1} (disregarding the $\pm 1$ cycle day$^{-1}$ alias components) have $6.2$, $5.5$ and $4.6$ $S/N$ ratio according to the definition given in \citet[Eq. 1]{al00}. The detected positions of the $f_0 - 2f_m$, $2f_0 - 2f_m$ and $3f_0 - 2f_m$ frequencies in this spectrum are very close to the exact quintuplet positions with $-0.000004, -0.00002$ and $0.00007$ cycle day$^{-1}$ separations.  According to Monte Carlo simulations, the by chance appearance of 3 frequency components with similar $S/N$ ratio in such a close vicinity of the exact quintuplet positions has practically zero probability.

In order to prove the reality of the appearance of the $kf_0 \pm 2f_m$ frequencies, we have performed a comprehensive analysis of the $B1$, $V1$, $B2$, $V2$ data sets. The determination of the $f_0$ and $f_m$ frequencies and the prewhitening follow three different procedures.

$a$) The $f_0$, $f_m$ frequencies are defined as the frequencies that give the best  light curve solution with the smallest rms residual. This solution takes into account altogether 36 frequencies simultaneously (pulsation components up to the 13th order, modulation $\pm$~side components up to the 11th order and the modulation frequency $f_m$). The data sets are prewhitened in one step with all the
pulsation and equidistant modulation frequency components.

$b$) The pulsation and modulation frequencies are determined separately.
The pulsation frequency is defined as the frequency which gives the smallest 
rms of the folded, modulated light curve. The modulation frequency is derived twofold,
first as the frequency which fits the maximum brightness variation the best,
secondly, modulation frequency components are searched in the prewhitened spectrum,
and the mean value of their separation from the pulsation components is accepted as $f_m$.
These two methods, however, give the same values for $f_m$ within the error ranges.
The data sets are prewhitened in consecutive steps, first the pulsation frequency and its harmonics, then the equidistant modulation components are removed.

$c$) The frequencies used in this procedure are the same as in case $b)$, but the data are prewhitened with the pulsation and modulation triplet frequencies in one step.

\begin{figure}
\epsscale{.80}
\includegraphics[angle=0,width=16cm]{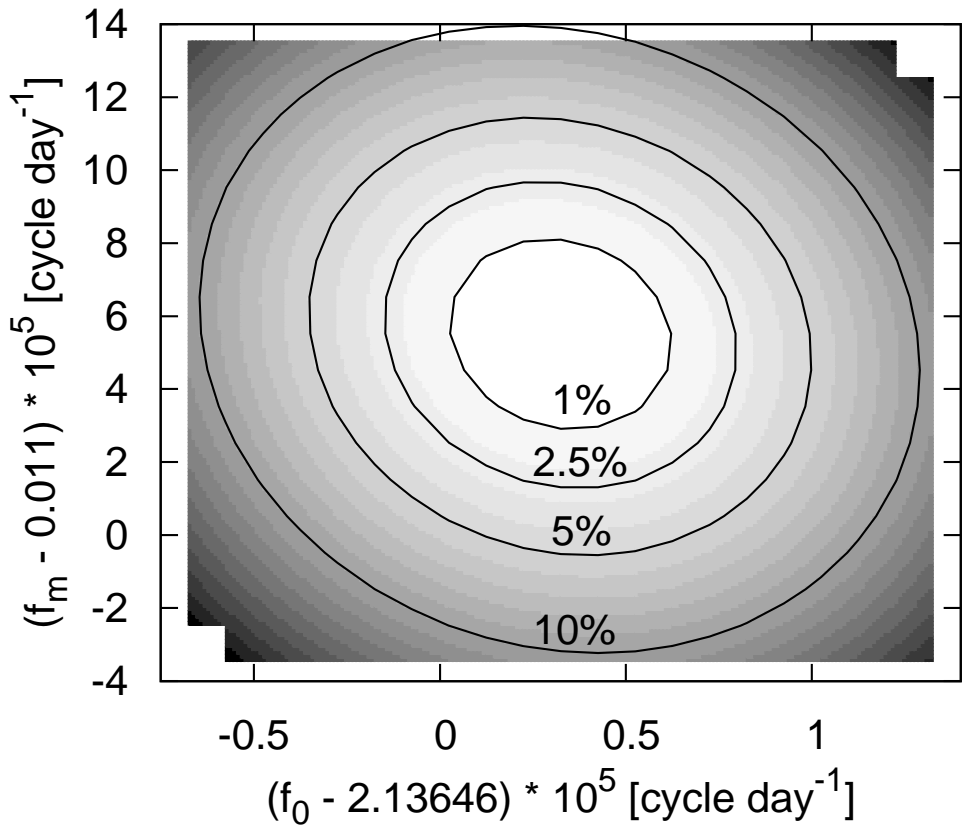}
\caption{
Contour map of the residual scatter of the $V2$ light curve fitted with 
equidistant triplet frequencies. The contours correspond to $1$, $2.5$, $5$, and $10\%$ increase in the residuals.
\label{res}}
\end{figure}

The $f_m$, $f_0$ frequencies determined for the $B1$, $V1$, $B2$ and $V2$ data sets according to methods $a)$ and $b)$ are listed in Table \ref{fqt}. In method $a)$ the errors of the $f_0$ and $f_m$ frequencies correspond to $10\%$ and $5\%$ increase of the residual scatter in the $V1$, $B1$ and $V2$, $B2$ data sets prewhitened with the triplet frequency solution, respectively. (As an example, Figure \ref{res} shows the residual scatter of the $V2$ data set in the $f_m$, $f_0$ plane.)
In method $b)$ the errors of the modulation frequencies are the $1 \sigma$ error of the sine wave fit to the maximum brightness data, while the errors of the pulsation frequency of the $V1$, $B1$ and $B2$, $V2$ data sets correspond to $0.025$ and $0.010$ phase shifts in the pulsation period during the time span of the observations. 
These error ranges were also checked visually as showing already noticeable differences in the
folded light curves: especially the disintegration of the fix point on 
the rising branch warns if we are out of the possible range of the pulsation period.
The less strict error estimate of the $B1$ and $V1$ data sets takes into
account that these data span over 17 years, and period changes had also a 
non-negligible effect on the residual scatters and phase incoherency 
during this time.

It has also been tested how accurately the positions of the $kf_0\pm f_m$  modulation frequencies are indeed equidistantly separated from the pulsation frequency components. Light curve solutions were calculated similarly as in 
case $a)$, but the modulation frequency $f_m$ was not fixed, the frequencies of the modulation components were
`let free' around their exact equidistant positions. The detected positions of the modulation components in these solutions deviated by $<\delta f_m> = 0.000019, 0.000003, -0.000004, 0.000003$ cycle day$^{-1}$ (error weighted means of the displacements of the frequencies of the $kf_0\pm f_m; k=1...9$ modulation components) from the exact equidistant positions for the $V1$, $B1$, $V2$, $B2$ data sets, respectively.
The displacements of the individual modulation components are within the ranges of
$-0.00023, +0.00037$ cycle day$^{-1}$ and $-0.00030, +0.00018$ cycle day$^{-1}$ for the $B2$ and $V2$ data sets.
The reality of the larger displacement of some of the modulation components is, however, questionable as none of the frequencies is simultaneously significantly shifted both in the $B$ and $V$ data.

\begin{figure}
\epsscale{.80}
\includegraphics[angle=0,width=16cm]{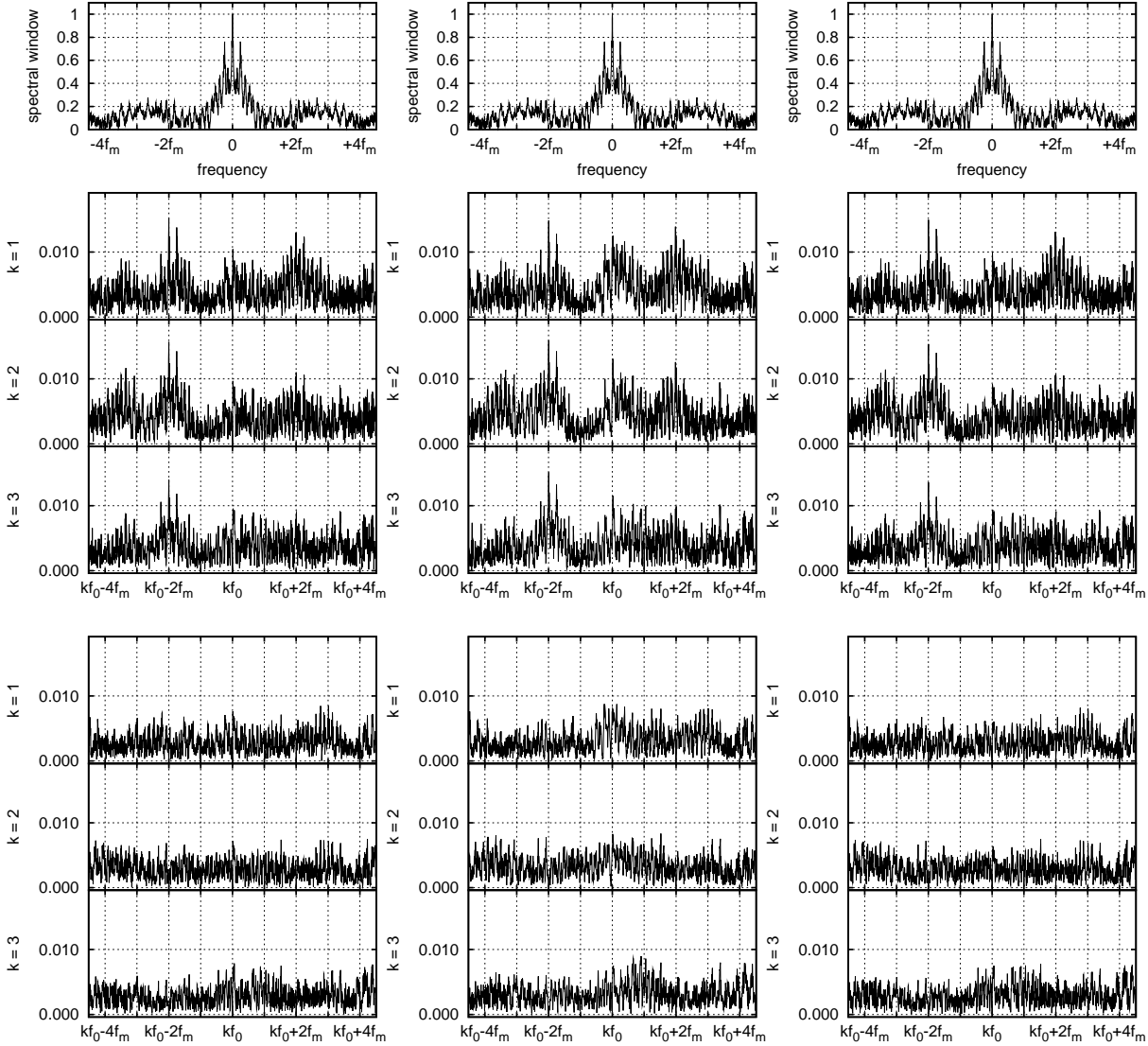}
\caption{Residual amplitude spectra [mag] {\it vs.} frequency [cycle day$^{-1}$] of the $V1$ data set in the vicinity of the $kf_0\ (k=1,2,3)$ pulsation frequency components. Top panels show the spectral window. Middle panels show the residuals after the removal of the pulsation and modulation components (triplet frequency solution). Panels in the left, center, and right show the residuals if prewhitening is done following the $a,b$ and $c$ procedures as described in the text. Bottom panels show the residuals if quintuplet frequencies are removed. Again, the prewhitening follows as described in the text. In the middle panels, independently from the method used to define $f_0$ and $f_m$ and from the prewhitening procedure applied, peaks at $kf_0 \pm 2f_m$ frequencies appear, which prove the reality of the appearance of quintuplet frequencies in the light curve of RV UMa. \label{vaf}}
\end{figure}

\begin{figure}
\epsscale{.80}
\includegraphics[angle=0,width=16cm]{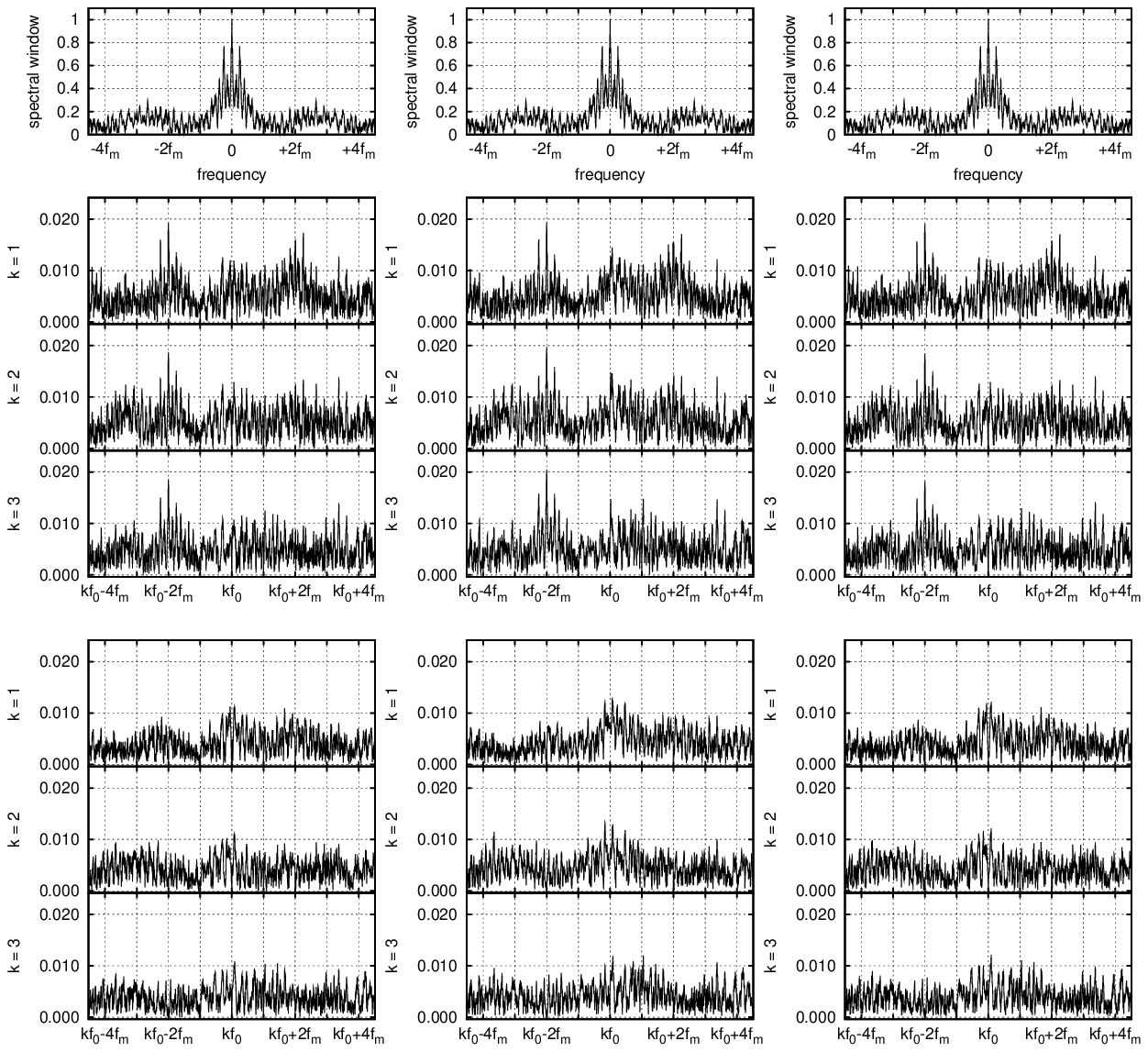}
\caption{The same as in Figure \ref{vaf} for the $B1$ data set.\label{baf}}
\end{figure}

\begin{figure}
\epsscale{.80}
\includegraphics[angle=0,width=16cm]{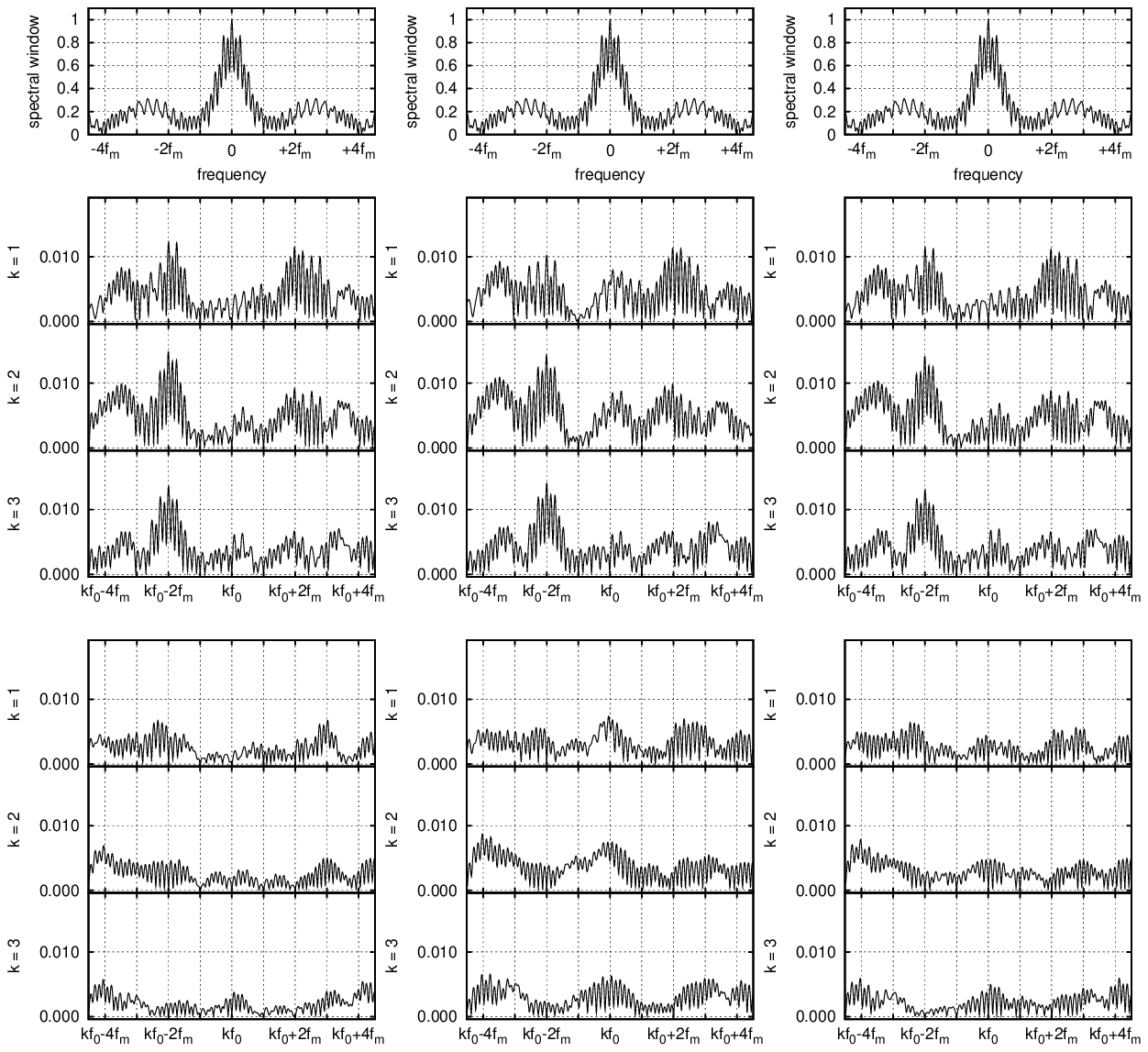}
\caption{The same as in Figure \ref{vaf} for the $V2$ data set.\label{vbf}}
\end{figure}

\begin{figure}
\epsscale{.80}
\includegraphics[angle=0,width=16cm]{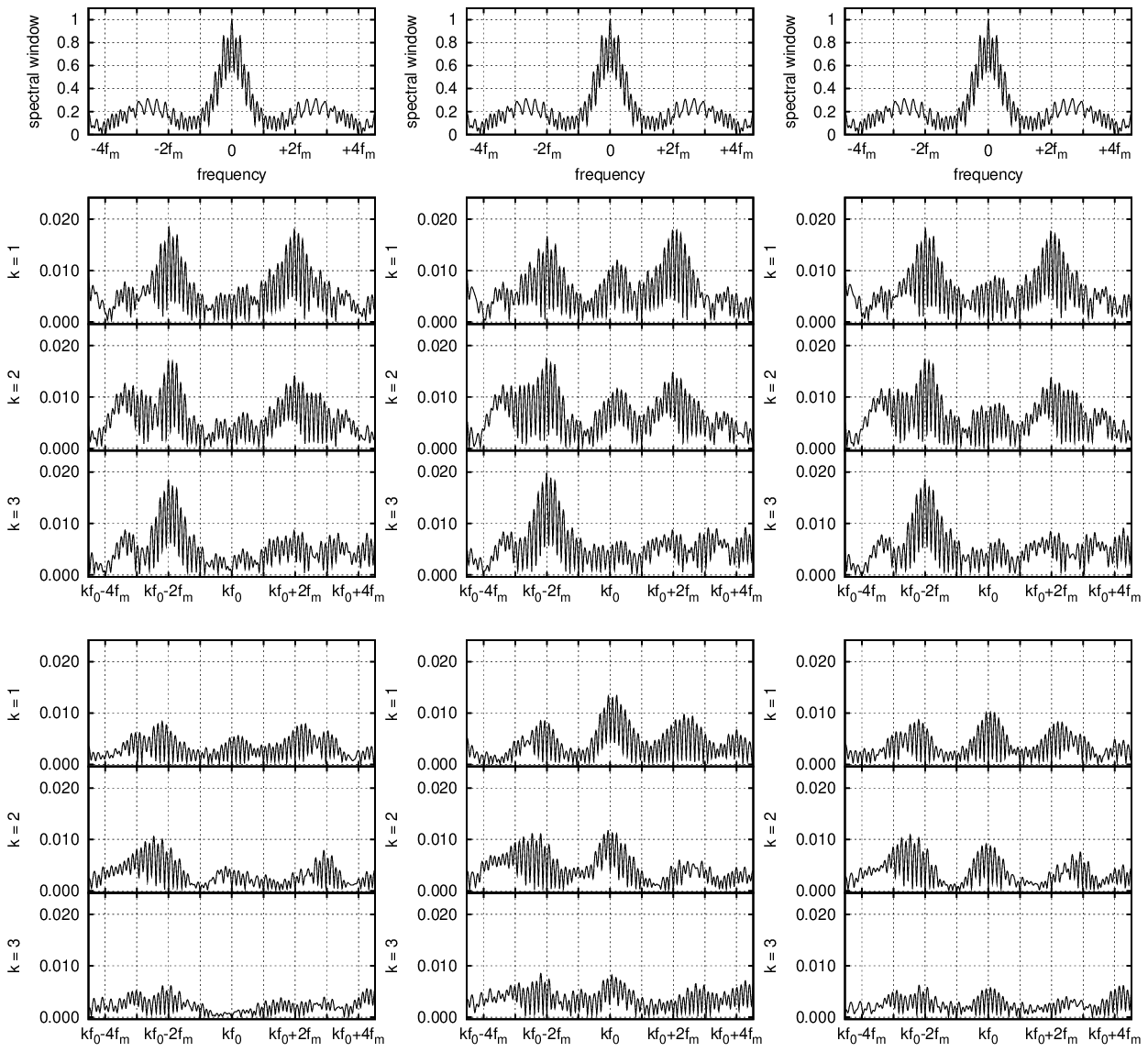}
\caption{ The same as in Figure \ref{vaf} for the $B2$ data set.\label{bbf}}
\end{figure}

Figures \ref{vaf}--\ref{bbf} show the residual spectra in the vicinity of the pulsation components for the $V1$, $B1$, $V2$, $B2$ data sets, respectively. The top panels are the spectral windows. The next panels display the residual spectra in the vicinity of the $f_0$, $2f_0$ and $3f_0$ frequencies. For the sake of clarity the separation of the tick marks on the frequency axes equals $f_m$ in all the figures. The residual spectra shown in the left, middle and right panels are calculated following the methods $a$, $b$ and $c$, respectively, using locked frequency values.
The `let free' solutions give the same results, its effect on the residual spectra is even smaller than the differences between the residuals shown in the three panels in 
Figures \ref{vaf}--\ref{bbf}.

\subsection{Detection of quintuplet frequencies}

The striking feature of all the residual spectra is the appearance of `secondary' modulation frequency components at $kf_0 \pm 2f_m$ frequencies. These frequency components are evidently present both in the whole data sets and in the subsamples of the $B$ and $V$ data. The secondary modulation peaks occur in each plot in Figures \ref{vaf}--\ref{bbf}, their appearance is independent from the method how the  $f_0$, and  $f_m$ frequencies have been determined and
whether the prewhitening has been done simultaneously with all the triplet frequencies or in consecutive steps.
Therefore, we conclude that the $kf_0 \pm 2f_m$ frequency components
are unambiguously inherent to the data, and they are not the manifestation of any defect (e.g., period change) of the data and/or the analysis.

The actual positions of the quintuplet components in `let free'
solutions match their exact equidistant positions within $0.00002 - 0.001$ cycle day$^{-1}$ without any systematic behavior.
E.g., the displacement of the $2f_0 - 2f_m$ component is $0.00032$ in 
the $B2$ data if a complete `let free' solution is applied with unlocked 
frequencies of all the triplet and quintuplet components, the exact 
position of this frequency component is displaced by $-0.0005$ for the 
$V2$ data prewhitened with the triplet frequency solution of Table \ref{fat} and 
its displacement is $0.00007$ if the data set is prewhitened  using frequencies listed 
in Table \ref{fqt}.

\begin{figure}
\epsscale{.80}
\includegraphics[angle=0,width=16cm]{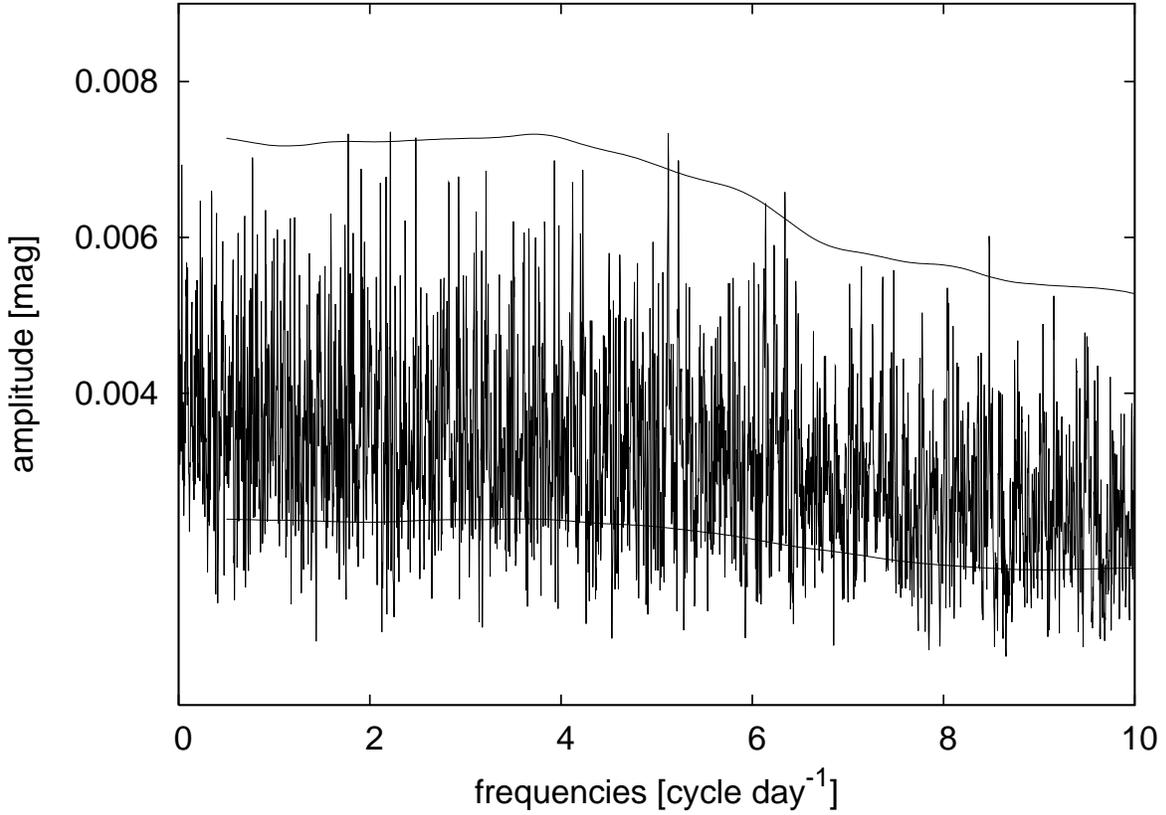}
\caption{Residual spectrum of the $V2$ data set after the removal of the equidistant quintuplet frequency solution from the light curve. The mean and the 4$\sigma$ level of the spectrum are also drawn.
The residual signals have amplitudes smaller than $0.007$ mag. The signals in this spectrum are already at the noise level compared to the spectrum shown in Figure \ref{v2res1}. None of the peaks higher than 4$\sigma$  appear  in the residual of the $V1$ data set, indicating that they cannot be identified with real signals. Compressed plotting is applied. \label{v2res2}}
\end{figure}

The bottom panels in Figures \ref{vaf}-\ref{bbf} show the residual spectra in the vicinity of the pulsation frequency components prewhitened with the $kf_0 \pm 2f_m$ frequency components as well. In the left panels, $f_0$ and $f_m$ have been recalculated assuming quintuplet frequency pattern, and altogether 13 pulsation and 29 modulation frequencies are simultaneously removed. In the middle panels data are prewhitened first with the pulsation, then with the `first order' ($kf_0 \pm f_m$) modulation, and finally with the `second order' ($kf_0 \pm 2f_m$) modulation components. Frequencies are those defined in method $b$. In the right panels the same frequencies are used as in the middle panels but, again, all the frequencies of the quintuplets are simultaneously fitted and removed. The amplitudes in the residual spectra are smaller than $0.008$ mag and $0.012$ mag for the $V$ and $B$ light curves, respectively, without any further common frequency component in each of the residuals. Figure \ref{v2res2} shows the residual spectrum of the $V2$ data set in the $0-10$ cycle day$^{-1}$ frequency domain. In this spectrum there is no frequency peak which would simultaneously appear in the residuals of the other data sets.

It is important to note that the residuals in the  $a$, $b$, and $c$ cases are not identical due to the procedures applied and the small differences in the frequencies used. The residuals are the smallest when the pulsation and modulation frequencies are simultaneously fitted and removed (methods $a$ and $c$). If the prewhitening is performed in consecutive steps (method $b$), the residual is larger, especially in the close vicinity of the pulsation frequencies. These residuals arise from the incorrect shape of the removed mean pulsation light curve due to biases in data sampling (e.g., the mean light curve has smaller amplitude than the real one if the small amplitude phase of the modulation is oversampled in the data). This experience warns that if the modulation has commensurable amplitude to the amplitude of the pulsation, then the best light curve solution can be yielded when the pulsation and modulation frequencies are simultaneously fitted, instead of removing first the mean pulsation light curve and then fitting the modulation components to the residual light curve.

As the $B1$ and $V1$ data sets are somewhat affected by the period changes the Fourier decompositions of the $B2$ and $V2$ subsets give amplitudes and phases more reliably. Table \ref{fat} lists the amplitudes and phases of the detected frequencies of  the $B2$ and $V2$ data sets assuming equidistant quintuplet frequency solutions. The $f_0$, $f_m$ frequencies are the averages of the frequencies which give the best fit to the $B2$, $V2$ light curves with the quintuplet frequencies involving 42 components ($kf_0$, $k=1...13$; $kf_0 \pm f_m$, $k=1...11$; $kf_0 \pm 2f_m$, $k=1...3$; $f_m$).
The differences between the frequencies defined 
for the two data sets are $0.000003$ and $0.00001$ for $f_0$, and $f_m$, respectively.
The results show agreement well within the different error estimates of the frequencies
given in Tables \ref{fqt} and \ref{fat}.
The residual scatter of the $B2$, $V2$ data are $0.030$ and $0.023$ mag, significantly smaller than the $0.060$ and $0.038$ mag rms that was given in \citet{kv95}.

The analysis of the completed and corrected Konkoly data supplemented with Preston \& Spinrad's observations of \object{RV UMa} led to the first definite detection of
equidistant quintuplet structure in the frequency spectrum of a Blazhko star. 
For this result both data sets are crucial, the quintuplet frequencies
are not evident if any of the observations are analyzed separately.
\citet{kv95} did not find any frequency component $kf_0 \pm jf_m$, $j>1$  in \object{RV UMa}, most probably because of the errors in the data sets he analyzed.
All the previous studies of other field Blazhko variables did not find quintuplet structure around the pulsation frequency components in their frequency spectrum either.
\citet{al04} analyzed 731 Blazhko RRab stars in the LMC but failed to find a quintuplet structure.
They gave an upper limit for the amplitude of the $f_0 \pm 2f_m$  components (if they exist at all) to be less than $0.004$ mag.
\citet{ch99} were not able to detect unambiguously quintuplet frequency components in the line profile variation of RR Lyrae's spectra. They explained it as ``{\it it is very likely that those frequencies have smaller amplitudes than the triplet components, they might not surpass the noise level in the periodograms of most of our frequency analyses}''. Whilst these studies estimated small amplitudes for the quintuplet frequency components vanishing below the noise level, we find e.g., for the amplitudes of the $2f_0 - 2f_m$ and $2f_0 + 2f_m$ components in the $V2$ data set $0.011$ mag and $0.006$ mag, respectively. Probably \object{RV UMa} is not a unique, peculiar Blazhko variable and the failure to detect quintuplet components in the frequency spectrum of other Blazhko RRab stars is simply due to the larger errors and/or the shorter time coverage of the observations or the smaller number of data in the data sets analyzed.

\section{The modulation frequency\label{mod}}

Using extended multicolor photometry of \object{RR Gem} and \object{SS Cnc}, it was shown in \citet{ju06} that the modulation frequency ($f_m$) has different properties compared to the pulsation and modulation side lobe frequencies. Especially the amplitude ratios of the modulation frequency in different colors were found to be anomalous. This might also be the case in \object{RV UMa}. While the mean value of the $A_B/A_V$ amplitude ratios of the first 9 pulsation components is $1.26 \pm 0.10$, and it is $1.25 \pm 0.06$  for the 10 largest amplitude  $kf_0 \pm f_m$ modulation components (the standard deviations correspond to
the rms of the amplitude ratios of the frequency components involved), the $A_B(f_m)/A_V(f_m)$ ratio is $1.44$. However, as the error of this amplitude ratio  (taking into account the strong correlation between the residuals in the $B$ and $V$ data) is
estimated to be about $0.1-0.2$, in RV UMa no definite conclusion about the anomalous behavior of the $A_B(f_m)/A_V(f_m)$ amplitude ratio can be drawn.

\citet{ju06} interpreted the anomalous behavior of the modulation frequency as the sign that the true physically meaningful frequency component in Blazhko stars is the modulation frequency itself and not the side frequencies. Further multicolor observations of other Blazhko variables are needed to strengthen the anomalous properties of the modulation frequency on a larger sample of Blazhko variables.

\section{Long-term period changes\label{ltc}}

\begin{figure}
\epsscale{.80}
\includegraphics[angle=0,width=16cm]{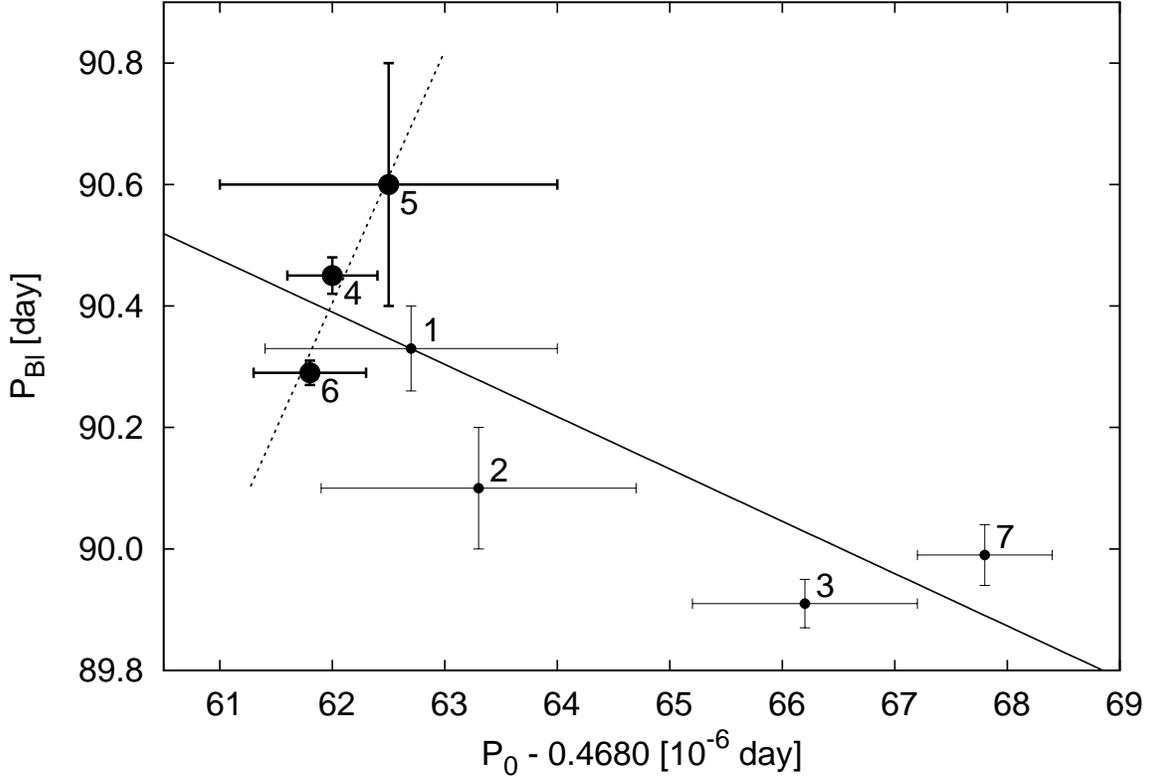}
\caption{Pulsation ($P_0$) {\it vs.} Blazhko ($P_{Bl}$) periods of RV UMa at different epochs between 1907 and 2000. The periods were determined using all the available photometric observations in \citet{hu07}. The numbers indicate the succession of the data in time, 1 is for the first observations from the beginning of the XX. century and 7
denotes the latest results from the Hipparcos and NSVS observations. On the average the period changes have the opposite sign, when the pulsation period is longer the modulation period is shorter and vica versa. The $dP_{Bl}$/$dP_0$ ratio defined by the period changes is $(-8.6 \pm 2.9) \times 10^4$. Between 1946 and 1975, however, parallel changes of the pulsation and modulation  periods were detected. Results for this time interval are denoted by larger symbols. The $dP_{Bl}$/$dP_0$ ratio for this period is $(4.2 \pm 1.1) \times 10^5$.\label{ppf}}
\end{figure}

Studying the long-term changes in the pulsation and modulation properties of Blazhko variables is an important issue that may help in understanding the modulation phenomenon. The recent finding that the modulation of \object{RR Gem} has changed from large amplitude phase modulation to amplitude modulation with small amplitude \citep{so07} suggests that
on longer time scales the properties of the modulation can change significantly. In the case of \object{RV UMa} there is no sign of any change in the amplitude of the modulation \citep{hu07}, however, due to data inhomogeneity it cannot be excluded that smaller changes might occur.
The 90 years long observations of \object{RV UMa} were collected and temporal pulsation and modulation periods were determined in \citet{hu07}. Detectable changes in the pulsation and modulation periods were observed, the pulsation and modulation periods varied within the period ranges of 0.468062 -- 0.468068 day and 89.9 -- 90.6 day, respectively. In Figure \ref{ppf} the pulsation {\it vs.} modulation period values determined for different epochs by \citet{hu07} are plotted. The two periods show anticorrelation with $dP_{Bl}$/$dP_0$=$(-8.6 \pm 2.9) \times 10^4$ gradient. Between 1946 and 1975 the two period changes had, however, the same sign (these data are shown with larger symbols in Figure \ref {ppf}). During this time interval the ratio of the period changes was positive: $dP_{Bl}$/$dP_0$=$(4.2 \pm 1.1) \times 10^5$.

\citet{la04} list four Blazhko stars with global $dP_{Bl}$/$dP_0$ gradients determined from $50-100$ years long observations. The detailed analysis of the long-term period changes revealed that the mean $dP_{Bl}$/$dP_0$ value of \object{RV UMa} is significantly smaller than given by \citet{la04}. The $dP_{Bl}$/$dP_0$ value of \object{RW Dra} is also revised (Szeidl 2007, private communication) its overall $dP_{Bl}$/$dP_0$ value is $-3.9 \times 10^3$. Now we can say that in all cases when the ratio of the period changes has been determined using long enough observations its value is in the order of $10^3-10^5$. Both positive and negative correlation between the period changes can be detected.

\section{Summary\label{sum}}

In this paper it is shown for the first time that the true mathematical model of the light curve of a Blazhko star, \object{RV UMa}, is a quintuplet instead of a triplet frequency solution. The detection of an equidistant quintuplet structure is of great importance in the explanation of the Blazhko phenomenon. \citet{sh00} showed that the light curve of an oblique magnetic pulsator has quintuplet frequency spectrum, while the nonradial resonant mode models \citep[e.g.,][]{vh98,dc99,nd01,dz04} would fail to explain the existence of the quintuplet frequencies. The lack of the $kf_0 \pm 2f_m$ frequency components was used previously as an argument against the oblique magnetic pulsator model. Though the detection of the quintuplet favors the oblique rotator explanation, we also have to keep in mind that \citet{ch04} recently failed to detect any magnetic field of \object{RR Lyrae}.

\acknowledgments

The authors would like to thank Katalin Ol\'ah for careful reading of the manuscript. Zs. H. thanks for the hospitality of the Konkoly Observatory. The constructive comments and suggestions of the anonymous referee helped us to improve the  paper significantly. The financial support of OTKA grants T-068626 and T-048961 is acknowledged.

\clearpage

\begin{deluxetable}{lcccc}
\tablewidth{0pt}
\tabletypesize{\scriptsize}
\tablecaption{Log of observations used in the analysis.\label{lot}}
\tablehead{
& \multicolumn{4}{c}{No. of nights / No. of data} \\
& \multicolumn{2}{c}{Konkoly observations} & \multicolumn{2}{c}{Preston \& Spinrad} \\
Year & \multicolumn{1}{c}{$B$} & \multicolumn{1}{c}{$V$} & \multicolumn{1}{c}{$B$} & 
\multicolumn{1}{c}{$V$} \\}
\startdata
$1958$ & $3$ / $69$ & $3$ / $69$ &  -- & -- \\
$1959$ & $6$ / $147$ & $6$ / $147$ & $11$ / $273$ & $11$ / $273$ \\
$1960$ & -- & -- & $5$ / $146$ & $5$ / $146$ \\
$1961$ & $31$ / $1711$ & $31$ / $1711$ & -- & -- \\
$1962$ & $13$ / $512$ & $13$ / $512$ & -- & -- \\
$1963$ & $8$ / $274$ & $1$ / $26$ & -- & -- \\
$1964$ & $5$ / $107$ & $5$ / $107$ & -- & -- \\
$1965$ & $2$ / $29$ & $2$ / $29$ & -- & -- \\
$1966$ & $2$ / $31$ & $2$ / $31$ & -- & -- \\
$1968$ & $2$ / $32$ & $2$ / $32$ & -- & -- \\
$1969$ & $2$ / $37$ & $2$ / $37$ & -- & -- \\
$1971$ & $8$ / $164$ & $8$ / $164$ & -- & -- \\
$1972$ & $1$ / $20$ & $1$ / $20$ & -- & -- \\
$1973$ & $5$ / $94$ & $5$ / $94$ & -- & -- \\
$1974$ & $2$ / $51$ & $2$ / $51$ & -- & -- \\
$1975$ & $1$ / $8$ & $1$ / $8$ & -- & --
\enddata

\end{deluxetable}

\begin{deluxetable}{lccllll}
\tablewidth{0pt}
\tabletypesize{\scriptsize}
\tablecaption{Summary of the four data sets analyzed.\label{fqt}}
\tablehead{
& & & \multicolumn{2}{c}{method $a$} & \multicolumn{2}{c}{method $b$} \\
Data set & JD & N & \multicolumn{1}{c}{$f_0$} & \multicolumn{1}{c}{$f_m$} & \multicolumn{1}{c}{$f_0$} & \multicolumn{1}{c}{$f_m$} \\
 & & & \multicolumn{2}{c}{[cycle day$^{-1}$]} & \multicolumn{2}{c}{[cycle day$^{-1}$]} \\
}
\startdata
$B1$ & $2,436,229-2,442,422$ & $3705$ & $2.136469(2)$ & $0.011064(19)$ & $2.136469(4)$ & $0.011067(3)$ \\
$V1$ & $2,436,229-2,442,422$ & $3457$ & $2.136469(2)$ & $0.011065(19)$ & $2.136470(4)$ & $0.011070(3)$ \\
$B2$ & $2,436,647-2,437,463$ & $2239$ & $2.136467(7)$ & $0.01105(6)$ & $2.136464(12)$ & $0.01103(3)$ \\
$V2$ & $2,436,647-2,437,463$ & $2239$ & $2.136463(7)$ & $0.01106(6)$ & $2.136461(12)$ & $0.01107(2)$ \\
\enddata
\end{deluxetable}

\begin{deluxetable}{lrlccrrccrr}
\tablewidth{0pt}
\tabletypesize{\scriptsize}
\tablecaption{Fourier amplitudes and phases of the quintuplet solutions for the $B2$, $V2$ data sets. Formal 1$\sigma$ errors of the amplitudes and phases are given. The errors of the frequencies are estimated from Monte Carlo simulations, derived as the scatter of the obtained frequencies in `let free' solutions of the simulated data of the $B2$ data set. Initial epoch $T_0$=$2,436,647.000$\label{fat}}
\tablehead{
& & & \multicolumn{4}{c}{$B$} & \multicolumn{4}{c}{$V$} \\
\multicolumn{2}{c}{frequency [cycle day$^{-1}$]}& err & $A$ [mag] & err & $\phi$\tablenotemark{*} [deg] & err & $A$ [mag] & err & $\phi$\tablenotemark{*} [deg] & err \\
}
\startdata
\multicolumn{3}{c}{$A_0$ / rms [mag]} & \multicolumn{4}{c}{$11.160$ / $0.030$} & \multicolumn{4}{c}{$10.852$ / $0.023$} \\
\multicolumn{11}{l}{Pulsation frequencies}\\
$f_0$ & $2.136466$ & $0.000003$ & $0.4956$ & $0.0012$ & $129.7$ & $0.1$ & $0.3661$ & $0.0009$ & $127.3$ & $0.1$ \\
$2f_0$ & $4.272933$ & $0.000007$ & $0.2155$ & $0.0014$ & $22.5$ & $0.3$ & $0.1686$ & $0.0011$ & $22.6$ & $0.3$ \\
$3f_0$ & $6.409399$ & $0.00001$ & $0.1250$ & $0.0013$ & $299.8$ & $0.6$ & $0.0952$ & $0.0010$ & $300.3$ & $0.6$ \\
$4f_0$ & $8.545865$ &  $0.00002$ &$0.0732$ & $0.0012$ & $211.3$ & $1.0$ & $0.0572$ & $0.0010$ & $210.4$ & $1.0$ \\
$5f_0$ & $10.682332$ & $0.00003$ & $0.0478$ & $0.0011$ & $121.2$ & $1.4$ & $0.0388$ & $0.0009$ & $127.0$ & $1.4$ \\
$6f_0$ & $12.818798$ & $0.00005$ & $0.0283$ & $0.0012$ & $34.7$ & $2.2$ & $0.0220$ & $0.0010$ & $30.0$ & $2.2$ \\
$7f_0$ & $14.955264$ & $0.00005$ & $0.0231$ & $0.0012$ & $303.9$ & $2.9$ & $0.0194$ & $0.0009$ & $301.8$ & $2.7$ \\
$8f_0$ & $17.091731$ & $0.00009$ & $0.0137$ & $0.0012$ & $229.8$ & $4.8$ & $0.0099$ & $0.0009$ & $226.1$ & $5.2$ \\
$9f_0$ & $19.228197$ & $0.00009$ & $0.0090$ & $0.0011$ & $124.8$ & $7.8$ & $0.0087$ & $0.0009$ & $134.1$ & $6.1$ \\
$10f_0$ & $21.364663$ & $0.0001$ & $0.0054$ & $0.0011$ & $6.0$ & $13.3$ & $0.0053$ & $0.0008$ & $3.7$ & $10.3$ \\
$11f_0$ & $23.501130$ & $0.0002$ & $0.0061$ & $0.0011$ & $315.6$ & $11.2$ & $0.0037$ & $0.0009$ & $319.4$ & $14.4$ \\
$12f_0$ & $25.637596$ & $0.0003$ & $0.0034$ & $0.0010$ & $271.5$ & $21.6$ & $0.0023$ & $0.0009$ & $234.4$ & $26.9$ \\
$13f_0$ & $27.774062$ & $0.0002$ & $0.0032$ & $0.0010$ & $133.3$ & $19.0$ & $0.0031$ & $0.0008$ & $138.0$ & $15.2$ \\
\multicolumn{11}{l}{Modulation triplet frequencies}\\
$f_m$ & $0.011034$ & $0.0002$ & $0.0078$ & $0.0013$ & $128.8$ & $10.8$ & $0.0054$ & $0.0011$ & $174.8$ & $11.4$ \\
$f_0 - f_m$ & $2.125433$ & $0.00002$ & $0.0646$ & $0.0016$ & $96.4$ & $1.2$ & $0.0476$ & $0.0012$ & $102.5$ & $1.3$ \\
$f_0 + f_m$ & $2.147500$ & $0.00002$ & $0.0602$ & $0.0013$ & $139.4$ & $1.3$ & $0.0475$ & $0.0010$ & $130.3$ & $1.3$ \\
$2f_0 - f_m$ & $4.261899$ & $0.00004$ & $0.0390$ & $0.0014$ & $331.4$ & $2.1$ & $0.0317$ & $0.0011$ & $328.4$ & $2.0$ \\
$2f_0 + f_m$ & $4.283966$ & $0.00003$ & $0.0540$ & $0.0013$ & $38.6$ & $1.6$ & $0.0418$ & $0.0010$ & $36.1$ & $1.6$ \\
$3f_0 - f_m$ & $6.398365$ & $0.00006$ & $0.0287$ & $0.0013$ & $225.4$ & $2.6$ & $0.0222$ & $0.0010$ & $223.0$ & $2.6$ \\
$3f_0 + f_m$ & $6.420433$ & $0.00003$ & $0.0349$ & $0.0014$ & $331.0$ & $2.4$ & $0.0277$ & $0.0011$ & $342.8$ & $2.3$ \\
$4f_0 - f_m$ & $8.534832$ & $0.00007$ & $0.0208$ & $0.0012$ & $169.3$ & $3.4$ & $0.0163$ & $0.0009$ & $166.4$ & $3.4$ \\
$4f_0 + f_m$ & $8.556899$ & $0.00004$ & $0.0216$ & $0.0012$ & $217.5$ & $3.5$ & $0.0193$ & $0.0009$ & $219.5$ & $3.1$ \\
$5f_0 - f_m$ & $10.671298$ & $0.00008$ & $0.0120$ & $0.0013$ & $74.4$ & $6.3$ & $0.0102$ & $0.0010$ & $93.3$ & $5.9$ \\
$5f_0 + f_m$ & $10.693365$ & $0.00007$ & $0.0158$ & $0.0012$ & $156.5$ & $4.4$ & $0.0127$ & $0.0010$ & $168.2$ & $4.2$ \\
$6f_0 - f_m$ & $12.807764$ & $0.0001$ & $0.0085$ & $0.0012$ & $34.1$ & $7.6$ & $0.0054$ & $0.0009$ & $25.4$ & $9.5$ \\
$6f_0 + f_m$ & $12.829832$ & $0.0001$ & $0.0098$ & $0.0012$ & $46.4$ & $7.0$ & $0.0071$ & $0.0009$ & $43.8$ & $7.6$ \\
$7f_0 - f_m$ & $14.944231$ & $0.0002$ & $0.0036$ & $0.0012$ & $289.6$ & $20.2$ & $0.0048$ & $0.0009$ & $300.7$ & $11.1$ \\
$7f_0 + f_m$ & $14.966298$ & $0.00006$ & $0.0125$ & $0.0012$ & $314.8$ & $5.2$ & $0.0120$ & $0.0009$ & $320.0$ & $4.2$ \\
$8f_0 - f_m$ & $17.080697$ & $0.0004$ & $0.0053$ & $0.0012$ & $202.0$ & $13.3$ & $0.0020$ & $0.0009$ & $206.3$ & $30.2$ \\
$8f_0 + f_m$ & $17.102764$ & $0.0002$ & $0.0068$ & $0.0012$ & $249.2$ & $9.5$ & $0.0045$ & $0.0009$ & $213.2$ & $11.7$ \\
$9f_0 - f_m$ & $19.217163$ & $0.0002$ & $0.0051$ & $0.0012$ & $170.1$ & $14.0$ & $0.0036$ & $0.0008$ & $175.5$ & $15.6$ \\
$9f_0 + f_m$ & $19.239231$ & $0.00009$ & $0.0071$ & $0.0012$ & $154.9$ & $9.8$ & $0.0079$ & $0.0009$ & $149.9$ & $6.9$ \\
$10f_0 - f_m$ & $21.353630$ & $0.0005$ & $0.0028$ & $0.0012$ & $83.0$ & $29.9$ & $0.0019$ & $0.0009$ & $327.9$ & $33.4$ \\
$10f_0 + f_m$ & $21.375697$ & $0.0002$ & $0.0046$ & $0.0011$ & $43.9$ & $15.4$ & $0.0041$ & $0.0009$ & $57.8$ & $13.5$ \\
$11f_0 - f_m$ & $23.490096$ & $0.0003$ & $0.0042$ & $0.0011$ & $358.4$ & $16.0$ & $0.0039$ & $0.0008$ & $332.8$ & $13.5$ \\
$11f_0 + f_m$ & $23.512164$ & $0.0002$ & $0.0040$ & $0.0010$ & $349.1$ & $16.7$ & $0.0040$ & $0.0008$ & $349.9$ & $12.9$ \\
\multicolumn{11}{l}{Modulation quintuplet frequencies}\\
$f_0 - 2f_m$ & $2.114399$ & $0.0002$ & $0.0113$ & $0.0014$ & $344.6$ & $6.2$ & $0.0051$ & $0.0010$ & $316.0$ & $12.1$ \\
$f_0 + 2f_m$ & $2.158534$ & $0.0001$ & $0.0112$ & $0.0012$ & $15.9$ & $7.8$ & $0.0072$ & $0.0010$ & $30.5$ & $9.4$ \\
$2f_0 - 2f_m$ & $4.250865$ & $0.0001$ & $0.0090$ & $0.0016$ & $164.0$ & $9.6$ & $0.0108$ & $0.0012$ & $174.6$ & $6.2$ \\
$2f_0 + 2f_m$ & $4.295000$ & $0.0001$ & $0.0070$ & $0.0014$ & $269.2$ & $13.2$ & $0.0055$ & $0.0011$ & $265.6$ & $13.1$ \\
$3f_0 - 2f_m$ & $6.387332$ & $0.0001$ & $0.0151$ & $0.0013$ & $121.7$ & $5.0$ & $0.0105$ & $0.0010$ & $127.4$ & $5.7$ \\
$3f_0 + 2f_m$ & $6.431466$ & $0.0005$ & $0.0024$ & $0.0012$ & $75.9$ & $36.7$ & $0.0021$ & $0.0009$ & $238.9$ & $34.0$
\enddata
\tablenotetext{*}{Phases correspond to sine decomposition.}
\end{deluxetable}

\clearpage


\clearpage

\end{document}